# Inflation Uncertainty, Output Growth Uncertainty and Macroeconomic Performance: Comparing Alternative Exchange Rate Regimes in Eastern Europe

Muhammad Khan[a]
khann.muhammad@gmail.com

Mazen Kebewar[a,b]
mazen.kebewar@univ-orleans.fr

Nikolay Nenovsky[a,c]
nenovsky@gmail.com

## Abstract

In the late 90's, after severe financial and economic crisis, accompanied by inflation and exchange rate instability, Eastern Europe emerged into two groups of countries with radically contrasting monetary regimes (Currency Boards and Inflation targeting). The task of our study is to compare econometrically the performance of these two regimes in terms of the relationship between inflation, output growth, nominal and real uncertainties from 2000 till now. In other words, we test the hypothesis of non-neutrality of monetary and exchange rate regimes with respect to these connections. In a whole, the empirical results do not allow us to judge which monetary regime is more appropriate and reasonable to assume. EU enlargement is one of the possible explanations for the numbing of the differences and the lack of coherence between the two regimes in terms of inflation, growth and their uncertainties.

[a] University of Orleans (France) – Faculty of Law, Economics and Management – Orleans Economic Laboratory (LEO), UMR CNRS 7322, Rue de Blois, B.P. 26739 – 45067 Orléans Cedex 2 – France.
[b] University of Aleppo (Syria), Faculty of Economics – Department of Statistics and Management Information Systems.
[c] CRIISEA, Université de Picardie Jules Verne, UFR d'Economie et de Gestion, Pôle Universitaire Cathédrale 10, Placette Lafleur, B.P. 2716 – 80 027 Amiens Cedex 1– France.

## 1. Introduction

In the late 90's, after severe financial and economic crisis, accompanied by inflation and exchange rate instability, Eastern Europe emerged into two groups of countries with radically contrasting monetary regimes. The first group was formed by the countries with Currency boards and strongly fixed exchange rate regimes (Estonia, Lithuania, Bulgaria and partly Latvia) and the second one was composed of Inflation targeting countries (Poland, Czech Republic, Hungary and later Romania). The reasons behind these choices were complex. One of leading arguments was the belief about the ability of two regimes to provide low inflation and to anchor inflationary expectations. This was viewed as a prepequisite for successful nominal and real convergence towards the EU integration and as a whole for sustainable economic growth. Over time, and especially with the launching of accession process and subsequently the EU membership, the differences in achievement and performance of both groups began to be increasingly subtle and unsystematic. This, in turn, undermined the importance of choosing one or another monetary regime and gave reasons to believe that the hypothesis of neutrality of the monetary regime can not be rejected.

The task of the present study is to compare econometrically the performance of these two regimes in terms of the relationship between inflation, inflation uncertainty, nominal and real uncertainty from 2000 till now. In other words, we test the hypothesis of non-neutrality of monetary and exchange rate regimes with respect to these connections. The article is constructed in three sections. The first section presents the theoretical foundations of the study, especially the main characteristics of both monetary regimes as well as the major theoretical relationships between nominal and real uncertainty. In the second section we set out and discuss empirical results. In the last section we conclude.

## 2. Theoretical framework

As already pointed out the choice of both polar regimes - Currency boards and Inflation targeting was dictated primarily by the necessity to curb inflation, to fix inflation expectations and to accelerate growth. In this line of reasoning credibility of the monetary regime, and its ability to establish discipline were the leading motives behind the choice of individual countries. Monetary regime is the primary institutional anchor that is systemic in nature, not only to inflation but also to the overall developments of the economy.

There is little doubt that both monetary regimes, namely Currency Board and Inflation targeting are very contrasting in their mechanics. The aim of Currency board is to import credibility and discipline from abroad by legally fixing exchange rate to the leading foreign currency and by means of full monetary base coverage with liquid foreign reserves. The monetary policy is eliminated because the balance sheet of central bank contains no domestic assets. The Central bank cannot perform open market operations, although some elements of monetary policy are available through the manipulation of reserve requirements and banking regulation. Currency Board relies on an automatic link between balance of payments and money supply, real exchange rate and interest rates are supposed to quickly address all imbalances. Balance sheet separation of the Treasury from the Central bank, obligates the government to pursue conservative fiscal policy, and as a rule to maintain fiscal surpluses and low public debt. Proponents of Currency board consider that it produces high levels of discipline and credibility.

Likewise, Inflation targeting pursues the same objectives (high credibility and discipline), but with others, and above all internal to the country mechanisms. These are clearly defined inflation target, transparency, as well as active conduct of monetary policy. It relies on good knowledge of the economic model and transmission mechanisms. In purely theoretical terms Inflation targeting requires fully floating exchange rate.[1] Supporters consider this monetary regime appropriate to combine the power of enhancing the level of fiscal discipline and credibility without eliminating the possibility for discretionary reaction in shocks.

Focusing on our sample countries we can say that Currency Board regimes are generally small and highly open peripherals economies pursuing quick integration into the monetary system of the developed European countries. For example, Currency boards in Estonia (1992) and Lithuania (1994), were introduced at the beginning of transition, the main objective explicitly was to break the influence of Russia and the Russian economy. Bulgaria, in turn, introduced Currency board in mid (1997) after deep financial and monetary crisis, period of hyperinflation and sharp devaluation of national currency.[2] Here the main task was to break with years of inflation, monetary instability and lack of structural reforms. In this sense, the choice of the Currency Board in Bulgaria can also be seen as a decisive geostrategic integration into the European monetary zone. Turning to the three countries with Inflation targeting (Poland, Czech Republic and Hungary, and to some extent Romania)[3], we see that they have the characteristics of Central European countries, they have some traditions in an economic and monetary policy prior to period of communism, and clear aspirations for independent and equal cooperation with leading European economies. Poland, the Czech Republic and Hungary began the transition with a fixed rate (to varying degrees) and progressively gained knowledge and experience in implementing monetary policy. These are countries that put much effort in building macroeconomic models serving the base for the later implementation of Inflation targeting regime. The case of Romania is somewhat peculiar. Romania has a number of characteristics similar both to Central European countries as well to Bulgaria that explains oscillations and late implementation of inflation targeting.

Turning to the theory, the multi-dimensional relationship between inflation, output growth as well as their uncertainties has been widely discussed in the literature.[4] Since a complete set of hypotheses is large, the empirical studies separately cover several aspects of this nexus. To begin with the inflation and nominal uncertainty relationship, Friedman's (1977) rule assumes a positive association between these two variables. Turning to the empirical literature, it mainly supports the Friedman's view of positive inflation effects on nominal uncertainty. As the survey results of Golob (1994) show that 17 out of 21 studies find a positive impact of inflation on nominal uncertainty.[5] Nevertheless, an alternative strand of literature argues that high inflation increases the cost of uncertainty and hence forces the agents to invest more time in predicting future prices (Frohman *et al.* 1981). Higher cost of ignorance, in terms of wealth and income loss, necessitates better information about these variables and hence high inflation becomes more predictable.

Concerning the feedback effect from uncertainty to inflation; two opposing sets of hypotheses have been forwarded by the literature. According to the first view that has been advanced by Cukierman and Meltzer (1986), higher uncertainty augments inflation by raising the short run benefits of inflation uncertainty. Notwithstanding the aversion of a long run higher inflation, policy maker seeks short run objectives of higher output from inflation surprises. This increases optimal inflation level in an economy. This positive relationship, which is known as 'Cukierman-Meltzer hypothesis' after the notation used by Grier and Perry (1998); is also supported by many empirical studies (see also Fountas *et al.* 2004). In sharp contrast to this view, Holland (1995) argues that central banks work for stability objectives, the so-called 'stabilizing Fed hypothesis'. As soon as uncertainty increases after inflation, central bank reacts by contracting money supply to avoid the welfare loss due to uncertainty; making this inflation and nominal uncertainty relationship negative. This hypothesis has been further complemented by empirical studies of Grier and Perry (2000) and Grier *et al.* (2004) etc. Finally, some mixed evidence based on country specific and time specific results are also documented by Caprole and Kontonikas (2009) among others.

---

[1] In practice the small and open economies such as in Eastern Europe still monitor and intervene on the foreign exchange market.
[2] Latvia carried out a similar policy of a fixed exchange rate; it is not the subject of our reflections notably because it does not represent Currency board in its pure form.
[3] Poland decided to move to inflation targeting in 1998 and introduce it in 2000, Hungary also passed in mid-2000, as well as Czech Republic - in 1997. Romania introduced this scheme in 2005.
[4] See Fountas and Karanasos (2007) for a brief survey of literature on all these aspects.
[5] See also Hess and Moris (1996) and Grier and Grier (2006), among others, for positive relationship between these two variables.

Regarding the inflation effects on real growth variability, literature usually complements the signalling extraction model of Lucas (1973). In this environment, inflation obscures the signalling channel of production and hence inhibits the output growth. Producers increase their production as an immediate response of price changes but as soon as they come to know the overall price change, they reverse back these decisions causing higher output growth variability due to inflation.

Likewise, inflation uncertainty also directly impacts output growth and its volatility. Nominal uncertainty influences growth through different channels; i.e through its effect on the long-term real interest rate, real wage, tax revenues, long-term investment plans of consumers and investors etc. Since the effects on all these channels are different and interdependent on the other policy organs, literature does not provide any precise and overwhelming evidence regarding the effect of uncertainty on output growth or its volatility. Pindyck (1991) substantiates an adverse effect of uncertainty on growth that appears through investment channel. Taylor (1979), on the other hand, assumes an inverse relationship between real and nominal uncertainties in the presence of real world rigidities. To illustrate, if supply shock hits the economy and real wages are rigid downward, real output fluctuations can only be avoided at the cost of higher nominal uncertainty. These results have been empirically supported by Cecchetti and Krause (2001), among others.

Finally, the relation between real growth volatility and growth has also been widely treated in the literature with (again) contradictory findings. First, according to Devereux's (1989) model, higher growth uncertainty, results lower degree of indexation. This further makes it easier for the policy maker to use inflation surprises as a tool for higher output objectives. This all ends up with high average inflation and lower growth. Second, according to Black (1987), uncertainty increases the degree of specialization in an economy and hence yields more growth. His work on business cycle fluctuations elaborate that higher degree of specialization in an economy will result a positive real growth and volatility nexus (see Fountas and Karanasos (2007) and references therein for empirical support on this view).

Regarding the selected sample of the present study, the new EU member states, we could point out again that all abovementioned theoretical links between inflation, growth and their uncertainties are carried out within the two fundamentally different monetary regimes (Currency Board and Inflation targeting), each of which strives to achieve both nominal and real stability as well as their predictability. Monetary regime by itself is an essential institutional environment which lays down the basic features of the whole set of behavioral relationships (Maurer, 2006).

## 3. Empirical analyses and interpretations

### 3.1 Methodology

Here we briefly discuss our selected exponential GARCH (E-GARCH) model that has been actualized to estimate the stochastic component of real and nominal uncertainties for our new EU members. This specific functional form has an advantage of taking into account the asymmetry in the inflation and uncertainty relationship and hence provides a better way to test the Friedman's view. Since according to Friedman, higher inflation exerts larger uncertainty, using a symmetric variance of the error term (the standard GARCH models) as measure of uncertainty is a poor approach to test this hypothesis (see Brunner and Hess, 1993 for more details). Following the traditional notations, let $\pi_t$ and $y_t$ represent the inflation and the output growth, respectively and define the residual vector $\varepsilon_t$ as $\varepsilon_t = (\varepsilon_{\pi t} \varepsilon_{yt})'$. A more specific form of the bivariate VAR (p) model can be described as;

$$\pi_t = a_0 + \sum_{i=1}^{p} a_i \pi_{t-i} + \sum_{i=1}^{p} \rho_i y_{t-i} + \eta_i i_{t-1} + \tau Oil_{t-1} + \varepsilon_{\pi t}, \qquad (1)$$

$$y_t = b_0 + \sum_{i=1}^{p} b_i y_{t-i} + \sum_{i=1}^{p} \delta_i \pi_{t-i} + \lambda y_{EU(t-1)} + \varepsilon_{yt}, \qquad (2)$$

Here '$i$' represents a change in the nominal interest rate. For the Currency Board economies we have used interest rate of European Central Bank (ECB). Here is one of the main difference between the both monetary regimes, in the case of Currency Boards because of the lack of own policy rate, and because of the mechanics of the regime, the ECB policy rate directly shape home market interest rate. 'Oil' in the first equation shows oil price changes as exogenous variable and '$y_{EU}$' represents European Union Industrial Production Index (IPI); to incorporate the effect of regional shocks on the domestic output. Our volatility estimates are based on the vector autoregressive (VAR) models and the specific number of lags is selected using Akiake Information Criteria (AIC) and Schwarz Information Criteria (SIC). Since E-GARCH specification models the logarithm of the conditional variance, it does not impose non-negativity constraint on the coefficients. Uncertainty in both variables is captured in the following way:

$$\log h_{\pi t} = \alpha_0 + \alpha_1 \log h_{\pi, t-1} + \beta \left| \frac{\varepsilon_{t-1}}{\sqrt{h_{t-1}}} \right| + \gamma \frac{\varepsilon_{t-1}}{\sqrt{h_{t-1}}} \qquad (3)$$

$$\log h_{yt} = \alpha_0 + \alpha_1 \log h_{y, t-1} + \beta \left| \frac{\varepsilon_{t-1}}{\sqrt{h_{t-1}}} \right| + \gamma \frac{\varepsilon_{t-1}}{\sqrt{h_{t-1}}} \qquad (4)$$

Here parameter '$\gamma$' estimates the asymmetry in the relationship. When '$\gamma$' is positive then positive inflation change causes more uncertainty than the negative one of the same magnitude. Our next step estimation consists of the Granger causality tests to see the exact direction of effects between real and nominal variables as well as their uncertainties.

### 3.2 Data

To test all these set of hypotheses, we use monthly data of six Eastern European economies over the period of 2000-01 to 2011-07, using International Financial Statistics (IFS, 2011). The selected sample includes three currency board economies; Bulgaria, Estonia, and Lithuania, and three inflation targeting countries Czech Republic, Romania, and Poland. For these economies, inflation is represented by the annualized monthly difference of the logarithm of the Consumer Price Index [$\pi_t = \log(CPI_t / CPI_{t-1}) \times 1200$] and real output growth is also measured accordingly [$y_t = \log(IPI_t / IPI_{t-1}) \times 1200$]. Our stationarity tests indicate that all the variables are non-stationary at their level and stationary at first difference, hence we use first difference of all variables in the analysis.[6] Summary statistics of both these variables are given in Table (1).

**Table (1)** Data Description

|  | Inflation ($\pi$) | | | Industrial Production Growth (y) | | |
|---|---|---|---|---|---|---|
|  | Mean | Standard Deviation | JB Normality | Mean | Standard Deviation | JB Normality |
| Bulgaria | 5.73 | 3.34 | 4.98[c] | 4.17 | 9.92 | 50.69[a] |
| Estonia | 4.14 | 2.85 | 2.42 | 5.78 | 14.08 | 75.29[a] |
| Lithuania | 3.01 | 3.33 | 17.08[a] | 5.83 | 11.23 | 11.22[a] |
| Czech Rep | 2.46 | 1.83 | 13.34[a] | 4.33 | 8.98 | 83.08[a] |
| Poland | 2.85 | 1.58 | 3.14 | 5.63 | 6.75 | 14.56[a] |
| Romania | 11.21 | 7.71 | 53.13[a] | 3.81 | 5.82 | 47.81[a] |

One prominent difference between inflation targeting and the currency board economies is the standard deviation of *y* and $\pi$ which is higher for the Currency Board economies than for the Inflation targeting countries. Since Inflation targeting regime explicitly focuses on the price stability, it reduces real uncertainty more efficiently than the Currency Board based monetary regime. Once nominal exchange rate is fixed under Currency Board, the uncertainty move to other variables especially price level. The case of Romania is distinctive. The inflation volatility is higher, because of a whole unstable economic and political climate.

### 3.3 Empirical Results

In order to apply our selected AR (p) - E-GARCH (1, 1) model we need to have the estimates of real and nominal uncertainty that are free from autocorrelation but contain heteroscedasticity; to justify the manipulation of this econometric technique. To do this, we ran AR (12) model and obtained the following results for the residuals.[7]

**Table (2)** Preliminary tests on the residuals

|  |  | Bulgaria | Estonia | Lithuania | Czech Rep | Poland | Romania |
|---|---|---|---|---|---|---|---|
| **Inflation** | $Q_{12}$ | 5.56 | 4.46 | 4.45 | 4.72 | 12.05 | 6.01 |
|  | $Q_1^2$ | 5.26[a] | 0.01 | 0.23 | 7.35[a] | 7.43[a] | 0.07 |
|  | $Q_{12}^2$ | 10.32 | 4.46 | 8.48 | 9.58 | 17.45 | 19.66[c] |
| **Output** | $Q_{12}$ | 7.32 | 4.82 | 9.11 | 8.25 | 6.91 | 11.33 |
|  | $Q_1^2$ | 4.01[b] | 0.33 | 4.45[b] | 5.74[a] | 1.39 | 3.17[c] |
|  | $Q_{12}^2$ | 10.68 | 26.71[a] | 10.43 | 31.12[a] | 25.12[a] | 23.77[a] |

Notes: $Q_{12}$ is the 12$^{th}$-order Ljung-Box test for standardized autocorrelation. $Q_1^2$ and $Q_{12}^2$ are tests for ARCH effects using squared residuals. a, b and c show 1%, 5% and 10% level of significance, respectively.

Our residual diagnostics (Table 2) show no sign of remaining autocorrelation for both inflation and output growth series. However the LM style ARCH tests show conditional heteroscedasticity in all cases except for two inflation series (Estonia and Lithuania). Although our square residuals are not so compelling in few cases, the results were, however, significant on some other lags; not mentioned here. Hence we use our AR (p) – E-GARCH (1,1) test for both inflation and output series and the results are presented in Table (3).

---

[6] We use Augmented Dickey Fuller (ADF) and Phillips-Peron (PP) tests for stationarity. The results are available on demand.

[7] Our AR (12) specification is based on the minimum values of AIC and SIC under alternative lags for all countries.

**Table (3)** Estimates of AR(p)-EGARCH(1,1) model of inflation

| | Bulgaria | Estonia | Lithuania | Czech Rp | Poland | Romania |
|---|---|---|---|---|---|---|
| Intercept $a_0$ | -0.025[a] | 0.001 | -0.017[a] | 0.021[a] | -0.005 | 0.006 |
| $a_1$ | 0.228[a] | 0.227[a] | 0.105 | -0.111[a] | 0.448[a] | 0.135[a] |
| $a_2$ | 0.124 | | | 0.152[b] | -0.255[a] | |
| $a_3$ | | 0.091 | 0.149[b] | | | |
| $a_4$ | -0.138[c] | | | | 0.19[c] | 0.116[c] |
| $a_5$ | | | | | -0.407[a] | |
| $a_6$ | -0.199[a] | | | -0.147[c] | 0.378[a] | 0.146[b] |
| $a_7$ | | | -0.106 | 0.053 | 0.378[a] | -0.137[c] |
| $a_8$ | | | -0.098 | | 0.233[b] | |
| $a_{10}$ | -0.128[c] | 0.093 | | -0.102[c] | -0.272[a] | -0.015 |
| $a_{11}$ | | -0.106 | | | 0.388[a] | |
| $a_{12}$ | -0.038 | 0.012 | 0.246[a] | 0.282[a] | | 0.008 |
| $\rho$ | 0.026[a] | 0.036 | 0.018[a] | 0.021[a] | 0.006 | -0.005 |
| $\eta$ | -0.0079[b] | -0.004 | -0.005 | 0.003[b] | 0.001 | -0.0003 |
| $\tau$ | | -0.0001[c] | | -0.0001[a] | -8.5E-05[c] | |
| $\alpha_0$ | -5.048[b] | -4.55 | -8.56 | -16.971[a] | 21.435[a] | -2.014[b] |
| $\alpha_1$ | 0.511[b] | 0.57 | 0.185 | -0.519[a] | -0.839[a] | 0.782[a] |
| $\beta$ | 0.318 | -0.19 | -0.406[b] | -0.661[b] | -0.295 | -0.52[a] |
| $\gamma$ | 0.615[a] | -0.17 | 0.128 | 0.523[a] | -0.081 | 0.116[b] |
| $R^2$ | 0.18 | 0.22 | 0.21 | 0.39 | 0.37 | 0.11 |
| F-Statistics | 1.84 | 2.12 | 2.11 | 3.78 | 3.58 | 0.88 |
| Log Likelihood | 397.96 | 558.61 | 459.09 | 440.25 | 514.06 | 464.52 |
| $Q_{12}$ | 7.817 | 7.92 | 4.63 | 5.31 | 11.61 | 8.14 |
| $Q_1^2$ | 0.01 | 0.04 | 0.41 | 0.01 | 0.65 | 0.19 |
| $Q_{12}^2$ | 5.36 | 4.88 | 7.98 | 10.27 | 10.59 | 12.09 |

Notes: $Q_{12}$ is the 12$^{th}$-order Ljung-Box test for standardized autocorrelation. $Q_1^2$ and $Q_{12}^2$ are tests for ARCH effects using squared residuals. a, b and c show 1%, 5% and 10% level of significance, respectively.

Here we present the results of inflation equations (1) and (3) for all countries. Most of the autoregressive coefficients are significant in almost all countries, showing strong influence of previous inflation to determine prices today. On our selected covariates, changes in previous period output amplify the level of inflation today. Effect of interest rate changes is different; it augments inflation when domestic interest rate increases (i.e in case of Czech Republic) and decreases inflation when ECB interest rate goes up (i.e in Bulgaria). One interpretation can be that higher interest rate in the ECB can result capital flight from this Currency Board based economy which further lowers reserve money, money supply and as a result inflation in this small open economy. However this systematically different effect is significant only in these two countries. Oil price increases lower the inflation in all cases where their impact turns out to be significant.

Turning to our volatility results, GARCH impact is significant in all cases where our first step residual tests have identified the presence of heterscedasticity in the inflation process. The most interesting results come from our asymmetric parameter '$\gamma$' (its sign is positive in all significant cases). When $\gamma>0$, this substantiates the Friedman and Ball effect regarding the impact of inflation volatility on its level. Higher inflation is indeed more volatile and hence more costly in terms of its impact on relative prices and output growth in an economy. Since welfare cost of inflation increases at its higher levels, price stability becomes an optimal choice for these economies; irrespective of the differences in the monetary policy regimes in these countries. These are the cases of Bulgaria, Lithuania, Czech Republic and Romania. When $\gamma<0$, meaning that higher inflation is negatively related with inflation volatility, as in the case of Estonia and Poland, the results seem intuitive. Estonian Currency

Board is extremely credible, has a long successful history, and this is probably the reason of the negative correlation, and its difference with the other two Currency Boards countries. Bulgarian Currency Board was introduced after deep crisis, in Lithuania there were always hesitations about the commitment to this monetary regime. Concerning the Polish case experts share the view that Central bank has established good reputation and this in some respect is confirmed by negative sign of '$\gamma$'.

**Table (4)** Estimates of AR(p)-EGARCH(1,1) model of Output Growth

| | Bulgaria | Estonia | Lithuania | Czech Rp | Poland | Romania |
|---|---|---|---|---|---|---|
| Intercept $b_0$ | 0.271$^a$ | -0.005 | 0.664$^a$ | 1.159$^a$ | 0.624$^a$ | 1.154$^a$ |
| $b_1$ | 0.133$^b$ | | 0.281$^a$ | 0.118 | 0.225$^c$ | 0.026 |
| $b_2$ | 0.324$^a$ | -0.101 | 0.217$^b$ | -0.022 | 0.196$^b$ | -0.094$^b$ |
| $b_3$ | 0.179$^a$ | 0.083 | | 0.034 | 0.181$^b$ | -0.061 |
| $b_4$ | | 0.084 | -0.293$^a$ | -0.063 | -0.045 | -0.021 |
| $b_5$ | | 0.377$^a$ | | -0.101 | -0.081 | -0.039 |
| $b_6$ | | | | | 0.139$^c$ | |
| $b_7$ | | | | | -0.202$^a$ | |
| $b_8$ | | | -0.188$^a$ | -0.141$^c$ | -0.023 | 0.049 |
| $b_9$ | | | 0.116$^c$ | -0.161$^c$ | -0.033 | -0.318$^a$ |
| $b_{10}$ | -0.147$^a$ | | | -0.356$^a$ | -0.153$^b$ | -0.149$^a$ |
| $b_{12}$ | 0.631$^a$ | 0.028 | 0.187$^a$ | 0.512$^a$ | 0.599$^a$ | 0.431$^a$ |
| $\delta$ | -0.325 | -0.014 | -0.721 | 1.442 | -2.039$^c$ | -0.761 |
| $\lambda$ | | 0.001$^a$ | -0.007$^a$ | -0.007$^a$ | -0.001 | -0.004$^a$ |
| $\alpha_0$ | -2.728$^a$ | -2.554$^b$ | -2.573 | -1.488 | -1.844 | -10.53$^a$ |
| $\alpha_1$ | 0.681$^a$ | 0.719$^a$ | 0.619$^b$ | 0.791$^a$ | 0.823$^a$ | -0.421$^b$ |
| $\beta$ | 0.812$^a$ | 0.367 | 0.459 | 0.178 | 0.734$^b$ | 1.11$^a$ |
| $\gamma$ | 0.303 | 0.652$^a$ | 0.137 | 0.223 | 0.249 | 0.125 |
| $R^2$ | 0.67 | 0.08 | 0.54 | 0.73 | 0.67 | 0.63 |
| F-Statistics | 14.59 | 0.874 | 10.22 | 14.11 | 8.52 | 9.56 |
| Log Likelihood | 204.69 | 298.72 | 168.38 | 205.86 | 236.62 | 225.44 |
| $Q_{12}$ | 12.06 | 5.24 | 8.94 | 10.97 | 11.79 | 6.85 |
| $Q_1^2$ | 0.66 | 0.11 | 0.38 | 1.46 | 0.48 | 0.11 |
| $Q_{12}^2$ | 13.37 | 15.06 | 14.88 | 8.68 | 8.75 | 11.79 |

Notes: $Q_{12}$ is the 12$^{th}$-order Ljung-Box test for standardized autocorrelation. $Q_1^2$ and $Q_{12}^2$ are tests for ARCH effects using squared residuals. a, b and c show 1%, 5% and 10% level of significance, respectively.

In the next step, we want to see how different monetary policy regimes impact the output growth or its variability in these emerging markets (growth equation 2 and 4). Here again we obtain standard results for the relationship between growth variability and overall output growth. One notable outcome appears from the fact that in almost all economies (other than Estonia, probably because of its large trade share with Russia), regional shocks (in our case EU) lower growth of these economies. Since EU is the largest trading partner of our selected economies, fluctuations in the industrial production of EU countries exert negative impact on the output growth of these economies. Parameter $\delta$, that shows trade off between inflation and growth volatility, show negative value (except for Czech Republic) in accordance to dominant theories. Asymmetry in the output growth and its uncertainty is again positive, albeit, insignificant in most of the cases.

Finally, in order to know the exact nature of the relationship and the direction of effects between real and nominal variables, we ran Granger causality tests (table 5). Various directions of effects have been tested for different variables; nevertheless, here we present some important channels through which real and nominal variables can influence each other. In most of the cases, inflation yields high nominal and real uncertainty; again supporting

Friedman and Ball hypothesis. For example according to panel *A* inflation causes inflation uncertainty in Romania, Bulgaria and Lithuania, but does not cause it in Estonia. Again the Estonian case is particular because of the high level credibility of Estonian currency board. Similar manifestation could be seen in panel C, when again in Estonian case the output growth reduces inflation uncertainty. Nominal uncertainty, however, does not lead to a significant real uncertainty. In the same way, output growth uncertainty aggravates the nominal uncertainty and real growth in almost all significant cases.

**Table (5)** Bivariate Granger-causality tests on the relationship between inflation and output growth as well as their uncertainties

| Optimal Lag | Bulgaria | Estonia | Lithuania | Czech Rp. | Poland | Romania |
|---|---|---|---|---|---|---|
| *Panel (A) $H_0$: Inflation does not Granger-cause inflation uncertainty* | | | | | | |
| 4 lags | <u>29.41$^a$</u>(+) | 43.62$^a$ (-) | **18.76$^a$** (+) | **9.89$^a$** (-) | **4.68$^a$** (+) | 11.07$^a$ (+) |
| 8 lags | **12.15$^a$** (+) | 23.37$^a$ (-) | 8.66$^a$ (+) | 4.92$^a$ (-) | 2.53$^b$ (+) | **7.29$^a$** (+) |
| 12 lags | 5.68$^a$ (+) | **17.57$^a$** (-) | 5.06$^a$ (+) | 4.54$^a$ (-) | 1.98$^b$ (+) | 4.49$^a$ (+) |
| *Panel (B) H0 : Inflation does not Granger-cause output growth uncertainty* | | | | | | |
| 4 lags | 0.26 | 1.21 | **2.65$^b$** (-) | <u>1.21</u> | **1.26** | 0.82 |
| 8 lags | 0.84 | 0.76 | 1.11 | 2.32$^b$ (+) | 1.48 | 0.56 |
| 12 lags | 0.92 | 0.86 | 0.78 | **2.48$^a$** (+) | 1.73$^b$ (+) | 0.61 |
| *Panel (C) H0 :Inflation uncertainty does not Granger-cause output growth uncertainty* | | | | | | |
| 4 lags | 0.06 | 0.02 | 0.55 | 3.05$^b$ (+) | 1.16 | 1.75 |
| 8 lags | 0.32 | 0.56 | 1.41 | 3.66$^a$ (+) | 0.93 | 1.11 |
| 12 lags | 0.27 | 0.57 | 1.42 | **7.58$^a$** (+) | 0.97 | 0.32 |
| *Panel (C) H0 :Output Growth does not Granger-cause inflation uncertainty* | | | | | | |
| 4 lags | 0.46 | **3.42$^a$** (-) | 1.83 | 0.83 | 0.98 | 0.48 |
| 8 lags | <u>3.01$^b$</u> (-) | 2.23$^a$ (-) | 1.48 | 0.37 | 1.07 | 0.58 |
| 12 lags | 1.52 | 1.92$^b$ (-) | 1.23 | 0.79 | 1.54 | 0.45 |
| *Panel (D) H0 :Output growth uncertainty does not Granger-cause inflation uncertainty* | | | | | | |
| 4 lags | 0.24 | 1.52 | <u>0.39</u> | 0.82 | **1.42** | 0.25 |
| 8 lags | 0.18 | 1.13 | **2.42$^b$** (+) | 0.58 | 1.98$^c$ (+) | 0.43 |
| 12 lags | 0.28 | 1.28 | 1.78$^b$ (+) | 0.57 | 2.54$^a$ (+) | 0.89 |
| *Panel (D) H0 : Output growth uncertainty does not Granger-cause output growth* | | | | | | |
| 4 lags | 2.72$^b$ (+) | **2.73$^b$** (-) | <u>**1.54**</u> | 11.51$^a$ (+) | 5.45$^a$ (+) | 3.74$^a$ (+) |
| 8 lags | 2.28$^b$ (+) | 1.54 | 1.81$^c$ (-) | 4.57$^a$ (+) | 1.65 | 2.09$^b$ (+) |
| 12 lags | **2.25$^b$** (+) | 1.43 | 1.09 | **2.31$^b$** (+) | **1.92$^b$** (+) | **1.87$^b$** (+) |

Notes: The numbers in the first column give the lag structure. Figures are F-statistics. A (+) (-) indicates that the sum of the lagged coefficients of the causing variable is (positive) (negative). a, b and c show 1%, 5% and 10% level of significance, respectively. The **bold** (<u>underlined</u>) numbers indicate the optimal lag length chosen by AIC (SIC).

## 4. Conclusion

On the whole, the empirical results do not allow judging which monetary regime is more appropriate and is reasonable to assume, as Jeffrey Frankel has said few years ago that all is about the specific assessments and depends upon the concrete characteristics of each economy. Evidence of the importance of specific historic background in determining the effectiveness of any monetary regime is clear, for example, in case of Estonia, where Currency Board fits well into the overall institutional setting and ideologically liberal climate. Or, Poland, where discretionary monetary policy is conducted by Central bank with a well established reputation.

But there is something else. The results of our study lead to consider the negotiation process, legal and normative convergence with the EU that occur after 2000, expectations dynamics of economic agents to the future EU integration, and subsequently EU membership itself, as major economic anchors, which largely absorbed the importance and specificity of internal anchors (monetary regime). EU enlargement is one of the possible explanations for numbing of the differences and the lack of distinction between the two regimes in terms of inflation, growth and their uncertainties.